\documentclass{aa}  
\usepackage{graphicx}
\usepackage{txfonts}
\def\deg{$^{\circ}$}
\def\lya{Lyman $\alpha$}

\begin{document}

\title{Interplanetary Lyman $\alpha$ line profiles:
  variations with solar activity cycle. }

\author{Eric Qu\'emerais \inst{1}, 
Rosine Lallement \inst{1},
Jean-Loup Bertaux \inst{1},
Dimitra Koutroumpa \inst{1},
John Clarke \inst{2},
Erkki Kyr\"ol\"a \inst{3},
\and
Walter Schmidt \inst{3}, }
\institute{ Service d'A\'eronomie, Verri\`eres le Buisson, France,
 \and
Boston University, Boston, USA, 
\and
Finnish Meteorological Institute, Helsinki, Finland}

\offprints{E. Qu\'emerais }

\date{March 9, 2006; April 10, 2006}

%% ------------------------------------------------------ %%
%
%%  ABSTRACT
%
%% ------------------------------------------------------ %%

  \abstract
  % context heading (optional)
  % {} leave it empty if necessary  
{}
 % aims heading (mandatory)
   {Interplanetary \lya\ line profiles are derived from the SWAN H cell data measurements. The measurements cover a 6-year
    period from solar minimum (1996) to after the solar maximum of 2001. This allows us to study the variations of the
    line profiles with solar activity.}
 % methods heading (mandatory)
   {These line profiles were used to derive line shifts and line widths in the interplanetary medium for 
   various angles of the LOS with the interstellar flow direction. The SWAN data results were then
   compared to an interplanetary background upwind spectrum obtained by STIS/HST in March 2001. }
 % results heading (mandatory)
   {We find that the LOS upwind velocity associated with the mean line shift of the IP \lya\ line varies from
    25.7 km/s to 21.4 km/s from solar minimum to solar maximum. Most of this change is linked with variations
   in the radiation pressure. LOS kinetic temperatures derived from IP line widths do not vary monotonically 
   with the upwind angle
   of the LOS. This is not compatible with calculations of IP line profiles based on hot model distributions
   of interplanetary hydrogen. We also find that the line profiles get narrower during solar maximum.}
 % conclusions heading (optional), leave it empty if necessary 
   {The results obtained on the line widths (LOS temperature) show that the IP line is composed of two
   components scattered by two hydrogen populations with different bulk velocities and temperature. This is a clear
   signature of the heliospheric interface on the line profiles seen at 1 AU from the sun. }


\authorrunning{QUEMERAIS ET AL.}

\titlerunning{IP LINE PROFILES}

\maketitle

\section{Introduction}

The interplanetary background has been used to study the interplanetary hydrogen distribution
since its discovery in the late 1960's (Thomas and Krassa 1970; 
Bertaux and Blamont 1970). Lists of previous experimental  studies of the interplanetary 
\lya\ background can be found  in Ajello et al. (1987, 1994) or Qu\'emerais et al. (1994). 

The SWAN instrument on the SOHO spacecraft (Bertaux et al., 1995) is able to study the IP \lya\
line profile using hydrogen absorption cells. These devices are used to scan the line profile
taking advantage of the variation of the Doppler shift between the cells and the interplanetary hydrogen 
as the spacecraft rotates on its orbit around the sun. Previous studies of the IP line profile
with H cells were made by the Mars 6 spacecraft (Bertaux et al., 1976)
the Prognoz 5/6 UV photometers (Bertaux et al. 1985)  and the ALAE instrument
on ATLAS-1 shuttle mission (Qu\'emerais et al. 1994).

IP \lya\ line profiles can also be obtained from high-resolution UV spectrometers in the Earth's orbit.
Clarke et al. (1995, 1998) report measurements made by the GHRS (Goddard High-Resolution Spectrograph)
instrument on Hubble Space Telescope.
More recently, new spectra were obtained from the STIS intrument that replaced GHRS on HST.

Studying  the IP \lya\ line profiles has proved to be of interest for studying the heliospheric 
interface because effects imprinted on the hydrogen distribution at a large distance from the sun are still 
visible on \lya\ line profiles seen at 1 AU. This is due to the fact that the flow of hydrogen in the 
heliosphere is collisionless.

Izmodenov et al. (2001) have presented model computations that predict how the hydrogen distribution 
is affected by the heliospheric interface. { The heliospheric interface is the boundary region
where the expanding solar plasma interacts with the interstellar plasma}. 
Using the derived hydrogen distributions, Qu\'emerais and 
Izmodenov (2002) calculated line profiles including a complete model of radiative transfer effects. 
The derived line shifts and
line widths were clearly modified by the heliospheric interface.

In this paper, we continue the work presented by Qu\'emerais et al. (1999) about the line profile 
reconstruction technique based on the SWAN H cell data. This early work was concentrated on the first 
year of data. Here we show results obtained from 6 one-year orbits of SOHO around the sun. Some of 
the details of the data processing are not repeated here
and the interested reader should refer to the original paper. 
Previous analyses of the SWAN H cell data were published by  Costa et al. (1999), Lallement et al. 
(2005), and Koutroumpa et al. (2005).

The first section briefly presents the line profile reconstruction technique and the SWAN data. The 
following sections give the results obtained for the line shifts and the line widths, as well as 
the variations during the solar cycle.
The last section presents a STIS/HST upwind line profile obtained in March 2001 and compares the 
derived line shift and line width with the SWAN results. 

\section{SWAN H Cell Data from 1996 to 2003}

%\subsection{Data Presentation}

The SOHO spacecraft was launched in December 1995 from Cape Canaveral.
The SWAN instrument started operation in January 1996 and has been 
active since then except for a few periods of time 
(June to October 1998 and January 1999).

Characteristics of the SWAN instrument are given 
in {Bertaux et al.} (1995). The SWAN instrument is a UV photometer
with a passband between 110 nm and 160 nm. The instrument is made of two 
identical units placed on opposite sides of the SOHO spacecraft 
(+Z and -Z sides). 
Each unit is equipped with a periscopic scanning mechanism that allows 
to point the field of view in any direction of the half sky facing the 
side the unit is attached on.
The instantaneous field of view is 5\deg\ by 5\deg\ divided in 25 pixels. Each pixel 
has a 1\deg\ by 1\deg\ field of view. Measurements are performed every 15 seconds, 
with at least 13 seconds of integration time to get a good signal-to-noise ratio. 

Each unit is equipped with a hydrogen absorption cell. This cell is placed on the 
photon path to the detector. The cell is filled with molecular hydrogen and has MgF$_2$ 
windows. A tungsten filament passes through the cell. When a current goes through the 
filament, H$_2$ is partially dissociated into atomic hydrogen, which creates a small 
cloud that can absorb Lyman $\alpha$ photons. Typical values for the optical thickness 
of the active cell are around 3 to 5.
Descriptions of the  observing programmes and of the various subsets of 
data are detailed by { Bertaux et al.} (1997).  
Here we concentrate on the  hydrogen absorption cell 
measurements obtained between June 1996 and June 2003.

The most common observation programme of SWAN is the so-called
full-sky observation, during which each sensor unit covers the complete
hemisphere that is on its side by moving a two-mirror 
periscope mechanism. One full-sky observation is performed in 
one day. The data obtained by both sensors are 
then combined into one image of the whole sky at \lya.
It must be noted that the areas of the sky viewed by each sensor
overlap, which enables us to compare both sensors on a regular
basis. SWAN performs these observations 4 times per week. 
During the first year, one of the four full-sky observations
made each week was made with cyclic activation of the hydrogen
cells. For these observations, we then obtained a full-sky image
at \lya\ as well as a full-sky image of the reduction factor,
{ which is defined below}.
The mechanism is kept fixed during both measurements, cell OFF 
and cell ON, before moving to another direction in the sky. 

The reduction factor $R$ used in what follows, is a dimensionless
quantity. It is the ratio of intensities measured in a given direction when 
the cell is on (absorption) and when it is off (no absorption).
If we consider an incoming intensity expressed as $I(\lambda_c)$
where $\lambda_c$ is the wavelength in the cell rest frame and if 
we note $T(\lambda_c)$ the transmission function inside the cell,
the reduction factor is defined by
\begin{equation}
R = \frac{I_{on} }{I_{off}} = 1 - A = \frac{\int_{-\infty}^{+\infty} ~I(\lambda_c)~T(\lambda_c)~d\lambda_c}
         {\int_{-\infty}^{+\infty} ~I(\lambda_c)~d\lambda_c} .
\end{equation}

Note that the quantity $A = 1 - R$ measures the absorbed fraction of the
incoming intensity. The transmission function inside the cell can be 
approximated with excellent accuracy for low optical thicknesses ($\tau < 10$) by
\begin{equation}
T( \lambda_c) = \exp\left( -\tau~e^{-x^2} \right),
\end{equation}
where the variable $x$ is defined as the normalized frequency by
\begin{equation}
x = \frac{ \nu_c - \nu_o }{\Delta\nu_d} =
-\left(\frac{ \lambda_c - \lambda_o }{\Delta\lambda_d} \right).
\end{equation}
The $\nu$ variable represents frequency, $\lambda$ wavelength.
Here, $\lambda_o$ is the wavelength of the \lya\ transition
at 1215.66 \AA.
The Doppler width of the cell, $\Delta\lambda_d$, is related 
to the temperature and the thermal velocity in the cell by
\begin{equation}
\Delta\lambda_d = \frac{\lambda_o}{c} \cdot Vth_{cell} = 
\frac{\lambda_o}{c} \sqrt{\frac{2 ~k ~T_{cell}}{m_H} }.
\end{equation}
The absorbing power of a hydrogen cell can be characterised
by a quantity called the equivalent width, $W_\lambda$,
in wavelength units.
\begin{equation}
W_\lambda = \int_{-\infty}^{+\infty} ~\left[1-T(\lambda_c)\right]~d\lambda_c .
\end{equation}
Typical maps of reduction factor values are shown in Bertaux et al. (1997).

Qu\'emerais et al. (1999) have shown how H cell measurements over
a 1-year period can be used to reconstruct the line profiles of the IP glow.
The same technique was applied here, and we refer readers to Sect. 3 of 
Qu\'emerais et al. (1999). We use the same terms and notations as in this paper.

The data used in this work cover a much longer period of time than our previous
work which was concentrated on the first year of SWAN data. To reconstruct the line profiles,
we need a full rotation of the Earth on its orbit. For each year we selected data 
from early June of the year until end of May of the following year. For instance,
the spectra labelled 2000 were derived from measurements starting in June 2000 and ending
in May 2001. The reason for this sampling was the lack of data between June 1998 and 
October 1998. As a consequence the year 1998 (June 1998 to May 1999) is missing. 

As done by Qu\'emerais et al. (1999), we use Eq. (5) to calibrate the absorbing power 
of the cell. Hydrogen cells are known to age with time. This may be due to ageing of the
tungsten filament itself or due to trapping of hydrogen by the glass or even to some leak.
This results in a decreasing optical thickness of the cell for a given filament current level. 
Over a 6-year period, both H cells have aged in very different ways. The H cell in the unit attached
to the -Z side of the spacecraft seems to have lost all absorbing power very rapidly in 1999.  
This suggests that the cell has lost its H$_2$ rather rapidly.  The H cell in the other unit (+Z side)
still retains most of its H$_2$ cloud as shown by the strong absorption still seen in the 2005 data. 
However, the optical thickness has decreased with time. This was calibrated by determining
the equivalent width of the cell (Eq. 5) from the data (see Sect. 4.1 of Qu\'emerais et al. 1999). 

The values for the equivalent width of the +Z hydrogen absorption cell are given in Table \ref{caliplus}.
Note that we use the transformation into velocity units given by Eq. (11) of Qu\'emerais et al. (1999). 
These values were used to determine the line profiles as presented in section 4.2 of Qu\'emerais et al. (1999).

\begin{table*}
\caption{ Equivalent Width of the H Cells \label{equiwid}} 
\begin{tabular}{c|cccccc}
Year (starts in June)   & 1996 & 1997 & 1999 & 2000 & 2001 & 2002 \\
Equivalent Width (km/s) &  5.5 $\pm$ 0.1 & 4.9 $\pm$ 0.1 & 3.68 $\pm$ 0.18 & 3.61 $\pm$ 0.18 & 3.50 $\pm$ 
0.12 & 3.26 $\pm$ 0.13 \\
Equivalent Width (m\AA) &  22.3 $\pm$ 0.4 & 19.9 $\pm$ 0.4 & 14.9 $\pm$ 0.7 & 14.6 $\pm$ 0.7 & 14.2 
$\pm$ 0.5 & 13.2 $\pm$ 0.5 \\
\end{tabular}
\label{caliplus}
\end{table*}

\section{Line-of-sight velocities}

This section presents the results found for each of the 6 orbits analysed. As mentioned earlier, each data set
starts in June and ends at the end of May of the following year. There is roughly one map per week, which means that
about 50 maps are used to derive a line profile in each direction of the sky. 
In 1996 and 1997, both H cells were active, so we have a full-sky velocity map for both orbits. However, due to the
leak in the -Z unit H cell, only the northern ecliptic hemisphere is available from 1999 to 2003.
We do not show individual plots of the line profiles. Examples are given in Fig. 5 to 7 of Qu\'emerais et al. (1999).

The line-of-sight (LOS hereafter) velocities shown here correspond to the mean Doppler shifts of the line profile 
expressed in terms of velocity in the solar rest frame.  If the line is not symmetrical, as in
Qu\'emerais and Izmodenov (2002) for example, then there is a small difference between the mean Doppler shift 
and the maximum of the line. 

Following Qu\'emerais et al. (1999), we express the relation between velocity $v$ projected on the LOS and 
the wavelength in the solar rest frame as

\begin{equation}
\lambda - \lambda_o = \frac{\lambda_o}{c} \left( \vec{V} \cdot
\vec{U} \right) = \lambda_o \left( \frac{v}{c} \right)
\label{lambtov}
\end{equation}
where $v$ is the projection of the atom velocity $\vec{V}$ on the line of
sight $\vec{U}$ . Note that the direction of the LOS is opposite
the direction of propagation of the photon.
In that case, the LOS velocity $<\!v\!>$ is given by

\begin{equation}
<\!v\!> = \frac{1}{I_{\mathrm off}} \int_{-\infty}^{+\infty}
v~I(v) ~dv = \frac{c}{\lambda_o} <\lambda-\lambda_o> .
\end{equation}

\subsection{Comparison of the first two orbits}

Qu\'emerais et al. (1999) show the velocity map derived from data obtained in 1996 and early 1997. 
This map corresponds to the minimum of activity of cycle 22. In the case of a flow with constant 
velocity, the velocity projected on the LOS is simply 
$V_\infty \cdot cos \theta$, where $\theta$ is the angle of the LOS with the upwind direction and
$V_\infty$ is the constant bulk velocity. The actual velocity maps observed by SWAN are more complex 
than this simple case.

\begin{figure}
\noindent\includegraphics[width=6.0cm,angle=90]{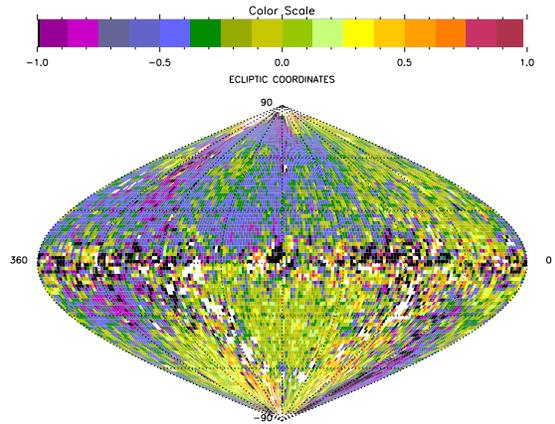}
\caption{ full-sky map of the velocity difference between the 1996 map and the 1997 map.
The data are shown in ecliptic coordinates (west ecliptic longitude). The blue color in the 
upwind side shows that the upwind velocity in 1997 is slightly slower (in modulus), by 0.4 km/s. }
\label{m96m97}
\end{figure}

Figure \ref{m96m97} displays the difference between the velocity maps of 1996 and 1997. 
The variation in the upwind velocity is a deceleration of 0.4 km/s, with a value of -25.7 km/s 
in 1996 and -25.3 km/s in 1997. Considering the uncertainties given in Table \ref{differvel}, 
the velocity can be considered to be the same.
However, a slight deceleration in the solar rest frame is quite possible because of the increase
in radiation pressure from the Sun between 1996 and 1997. The LOS (line-of-sight) velocity in the upwind 
direction is the result of two antagonistic effects. First, selection effects linked to ionization processes
increase the  mean velocity by ionizing the slower hydrogen atoms. Second, radiation pressure tends to push 
the atoms away from the sun and slow them down. During most of the solar cycle, radiation pressure is 
larger than the solar gravitational attraction (Pryor et al. 1998).
 As the cycle changes from minimum to maximum, the interplanetary
hydrogen atoms feel an increasing radiation pressure that becomes more efficient 
to slow them down in the upwind
side of the inner heliosphere.

\begin{figure}
\noindent\includegraphics[width=7.5cm]{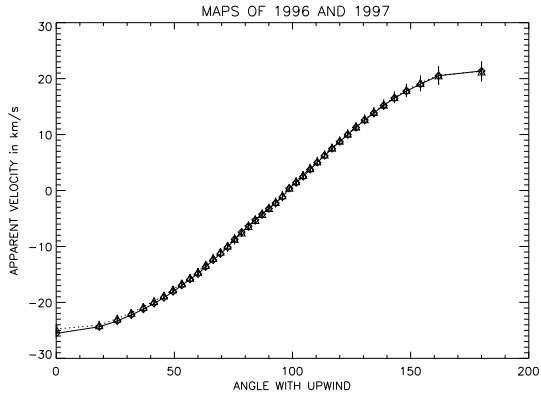}
\caption{ Variation in the LOS velocity as a function of the angle of the LOS
with the upwind direction. Here, the upwind direction is taken to be 252.3\deg , 8.7\deg . 
Both curves are very similar showing that the 1996 and 1997 velocity profiles are very close
and correspond to solar minimum conditions. The solid line joins the values found in 1996 and
the dotted line joins the 1997 values.}
\label{vel9697angle}
\end{figure}

The downwind velocity is mostly unchanged with a value equal to 21.6$\pm$1.3 km/s, which 
is not surprising because the hydrogen distribution in the downwind side of the heliosphere is 
less affected by variations in radiation pressure. 
In this direction, the hydrogen is strongly depleted by ionization effects. A partial filling happens
beyond a few AU from the sun  because of the relatively hot temperature of the interplanetary gas.
It is therefore not surprising that radiation pressure plays a lesser role in this direction as 
compared to the upwind direction.

Figure \ref{vel9697angle} shows the velocity in the solar rest frame as a function of the angle with the upwind direction.
The upwind direction is the one defined by the isovelocity contours (Qu\'emerais et al., 1999) with a longitude of
252.3\deg and a latitude of 8.7\deg . Changes between the two orbits are within the statistical uncertainties of the
velocity values.

\subsection{Variation from solar minimum to solar maximum}

The SWAN instrument did not operate between the end of June 1998 and October 1998, because of the accidental loss 
of contact with the SOHO spacecraft. Operations resumed at the end of 1998 but soon were stopped
after the failure of the last gyroscope. Normal operations resumed in spring 1999. 

In this section, we present maps obtained from measurements made beween June 1999 and June 2003. This period
covers the solar activity maximum of 2001 and the beginning of cycle 23.
There are no measurements in the southern ecliptic hemisphere because the -Z unit H cell stopped absorbing
in 1999 for a reason that is still unclear, while the +Z unit hydrogen cell is still functional in 2006. 
Table \ref{caliplus}
shows the variation in the equivalent width of the cell as a function of orbit year. Between 1997 and 1999, the +Z H cell
shows a strong decrease in absorption and after that a steady but slower decrease. The equivalent width in 2002 is only
60\% of the post launch value.

\begin{figure}
\noindent\includegraphics[width=7.5cm]{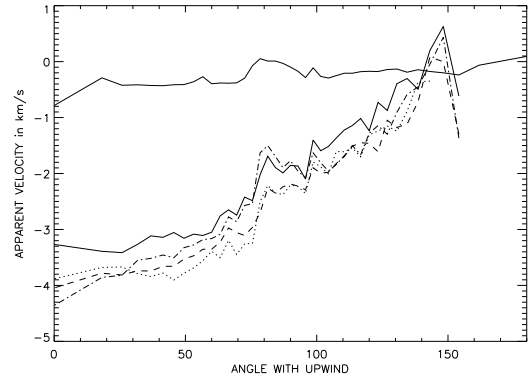}
\caption{ Difference in LOS velocity between  the 1996 orbit and the five other orbits. 
The velocities are shown as a function of the angle with the upwind direction. 
The 1996 velocity profile is shown in Fig. \ref{vel9697angle}.
The 1997 profile (solid line, top) is very close to the 1996 profile. The 
1999 one (solid line bottom) is an intermediary profile, while 
the 2000 (dotted line), 2001 (dashed line), and 2002 (dash-dot line) profiles are very similar.}
\label{vel9903angle}
\end{figure}

Figure \ref{vel9903angle} shows the difference between the 1996 orbit velocity profile and the five other orbits. 
The differences are shown as a function of the LOS angle from the upwind direction. First, we note that the 2000 
to 2002 orbits yield very similar values, within 1 km/s almost everywhere. Since the 1996 and 1997 orbits are also 
very similar, we see two main regimes for the velocity: 
one for solar minimum conditions with an upwind LOS velocity around -25.5 km/s and one for solar
maximum conditions with an upwind LOS velocity around -21.5 km/s. The 1999 and the missing 1998 ones
are intermediary states.  

\begin{table*}
\caption{ Upwind LOS velocity for various orbits \label{differvel}} 
\begin{tabular}{c|cccccc}
Year (starts in June)              &  1996 & 1997 & 1999 & 2000 & 2001 & 2002 \\
\hline
Upwind V(year) (km/s)   &  -25.7$\pm$0.2 & -25.3$\pm$0.2 & -22.5$\pm$0.5 & -21.5$\pm$1.2 & -21.5$\pm$0.3 & -21.4$\pm$0.5 \\
Upwind V(year) - V(1996)  (km/s)   & 0. &  0.4 & 3.2 & 4.2 & 4.2 & 4.3 \\
%Angle for V=0. & 90\deg &  90\deg & 90\deg & 90\deg & 90\deg & 90\deg \\
%Downwind V(year) - V(1996)  (km/s) & 1996 & 1997 & 1999 & 2000 & 2001 & 2002 \\
%Total V(year) - V(1996)  (km/s)    & 1996 & 1997 & 1999 & 2000 & 2001 & 2002 \\
\end{tabular}
\end{table*}

Table \ref{differvel} gives the numerical values of the upwind velocity for each orbit. The values are slightly different
from those in Fig. \ref{vel9903angle} because the averaging was done in a different way. In this table, velocities are
averaged within 20\deg\ of the upwind direction to get better statistics. The apparent variation in upwind velocity can 
be explained partially by effects of the radiation pressure; however, detailed calculations will be necessary to see if the
models match the measured variations.  The general deceleration observed in the solar rest frame
corresponds to increasing values of radiation pressure as expected from solar Lyman $\alpha$ flux measurements 
(Rottman et al. 2006). 
The change between 1997 and 1999 is very abrupt (3 km/s) and seems too important for the change of radiation pressure 
(+10\%) derived from  solar flux measurements.  
Changes of ionization processes could also be partly responsible for this change of LOS velocity.
Indeed, ionization processes favour fast atoms because slower atoms have a higher probability of being
ionized. This heads to the selection of fast atoms and an increase in the bulk velocity in the inner heliosphere. 
However even the fast atoms are ionized with increasing efficiency of the ionization processes.
This means that the atoms contributing to the intensity are farther away from the sun. How this will affect 
the line-of-sight velocity is not clear. Only  time-dependent calculation of the hydrogen distribution 
will allow us to discriminate between the various effects involved here. 

\begin{figure}
\noindent{\includegraphics[width=6.0cm,angle=90]{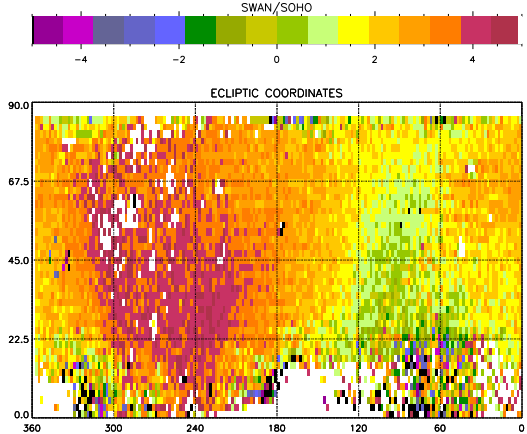}}
\caption{ Difference between the velocity maps of 2001 and 1996. The maximum difference is found in the upwind
direction with a value around 4 km/s. Downwind, the difference is close to zero. Below 10 degrees of latitude, 
the results are very noisy. Contours of the 1996 velocity map are shifted towards higher latitudes than the contours of the 
2001 velocity map. This is due to the existence of anisotropies in the hydrogen distribution of 1996 which create a deviation 
from the non-isotropic 2001 distribution.}
\label{fullmask}
\end{figure}

Figure \ref{fullmask} presents the LOS velocity difference between 2001 and 1996 for all directions in the northern ecliptic
hemisphere. The difference in the upwind direction is roughly 4 km/s, as seen before. The difference in the downwind direction 
is small, which suggests that changes in radiation pressure over the cycle is not as important with the velocity distribution
of hydrogen atoms in the downwind direction. We clearly see a shift between the iso-contours of velocity for
2001 and 1996. This is explained by the changes in the velocity contours induced by anisotropies of the solar ionization fluxes 
(Lallement et al. 2005; Koutroumpa et al. 2005). In 2001, when solar ionizing fluxes are almost isotropic, the constant velocity
contours are well fitted by cones centered on the upwind direction. In 1996, however, due to different ionization 
fluxes at different heliographic latitudes the resulting iso-velocity contours are elongated towards high latitudes. This is 
demonstrated in Fig. \ref{fullmask} where the difference of velocity maps shows a maximum that is
shifted towards higher ecliptic latitudes.

In this section,  we have shown how the LOS velocity of the interplanetary \lya\ line profile changes during the solar cycle.
The amplitude is large, more than 4 km/s in the upwind direction. It is also abrupt because we find
a change of 3 km/s between 1997  and 1999. The main cause for this variation in the velocity
is the change of radiation pressure during the solar cycle (Pryor et al. 1998). 
Other effects linked to ionization processes and their solar cycle variations may also be involved.

\section{Line-of-sight Temperatures}

This section presents the LOS kinetic temperature maps deduced from the line-profile reconstruction 
technique presented in Qu\'emerais et al. (1999). What we call the line-of-sight (LOS) kinetic temperature,
 also called apparent temperature in Qu\'emerais and Izmodenov (2002),  is actually the line width converted to
temperature units using the relation between thermal velocity $V_{\mathrm th}$ and line width.

Using the same notation as before, we have the following expression

\begin{equation}
T_{\mathrm ap} = \frac{m_{\mathrm H}}{2 k} ~V_{\mathrm th}^2
 = \frac{m_{\mathrm H}}{k} \int_{-\infty}^{+\infty} \!
\left(v-<\!v\!>\right)^2~\frac{I(v)}{I_{\mathrm off}} ~dv
\end{equation}
where $I(v)$ gives the intensity profile as a function of the LOS projected velocity $v$.
Line width values are more difficult to derive from the SWAN H cell data than line shifts. 
Indeed, if there is some stellar light, noted $I_{star}$ here,  
added to the \lya\ data for a given LOS,  the absorption becomes

\begin{equation}
A(measured) = 1 - R = \frac{ I_{off} - I_{on}} { I_{off} + I_{star}} = A(real) ~\frac{ 1 } { 1 + I_{star}/ I_{off}}.
\end{equation}

If the value of $I_{star}$ is not equal to zero, the value of the mean line shift will not be 
affected much, but it will change the integral of the absorption profile and thus change the 
derived temperature significantly.
To solve this problem, we used the measurements of the so-called BaF$_2$ pixel of the
+Z sensor. As mentioned by Bertaux et al. (1995), each detector has two active pixels on the
sides of the $5 \times 5$ array. One of the side pixels of the +Z sensor was covered
by a window made with BaF$_2$. This window is opaque for Lyman $\alpha$ photons. Using
the data recorded by this pixel over many months, we compiled a full-sky map 
that excludes the Lyman $\alpha$ background. This map can then be used to determine
which areas of the sky are free of stellar contamination. This way, we  created
a mask that allows us to keep only the data not contaminated by starlight.  

\begin{figure}
\noindent\includegraphics[width=7.5cm]{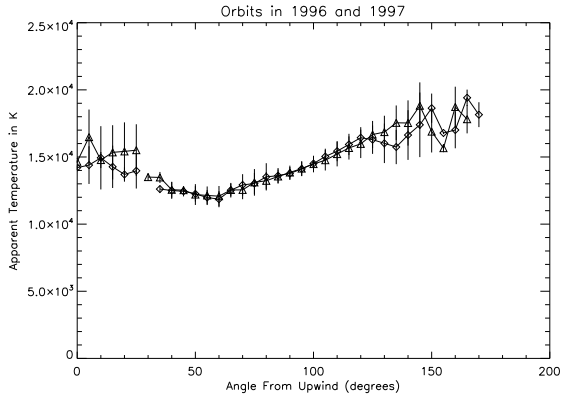}
\caption{ LOS kinetic temperature as a function of the upwind angle. The values were obtained for the two orbits
in 1996 and 1997. For each orbit, there are 2 curves, one for angles smaller than 30\deg\ and one for angles
larger than 30\deg. For angles smaller than 30\deg , the limit on stellar light counts has been relaxed to ensure
that some points with a small contamination are kept. Contrary to hot model predictions, the temperature curve is 
not monotonic and shows a minimum value of 11000 K for angles close to 60\deg . The 1996 values are given by the diamonds
and the 1997 values are given by the triangles. }
\label{temp9697}
\end{figure}

Figure \ref{temp9697} shows the LOS temperature as a function of the upwind angle derived from
the orbits of 1996 and 1997. 
For each orbit, there are 2 curves, one for angles larger than 30\deg\ and one for angles smaller
than 30\deg . Because the upwind direction is very close to the galactic plane, applying a strict
limit on stellar counts for each LOS removes all points within 30\deg\ from upwind.  
However, some lines of sight show a low stellar contamination, of a few percent of the upwind intensity.
Keeping those points, we were able to determine a curve for angles between zero and 30\deg .
This second curve shows larger uncertainties. The most striking feature of these two curves is that
they are not monotonic.  The temperature reaches a minimum value around 11000 K for an upwind angle of 60\deg .
This feature does not appear in hot model calculations whether they include full radiative transfer 
effects on the line profile or simply compute the first-order scattering of photons (Qu\'emerais 2000). 
For a hot model, the temperature dependence of line profiles from upwind to downwind is always monotonic. 
%({\bf \`a v\'erifier}).
 We also find that the upwind LOS temperature is around 14000 K and the downwind
value is close to 18000 K. This departure from a monotonic variation with the upwind angle was also 
noted by Costa et al. (1999) using a different method. 

Qu\'emerais (2000) computed LOS temperatures as seen from Earth's orbit for various models
of the hydrogen distribution. What was found is that the LOS temperature of the IP line within 
40\deg\ to 50\deg\ in the upwind direction is almost constant. The line width starts to increase for angles larger
than 50\deg . This increase in the line width is due to changes of the line shape due to acceleration and 
deceleration by radiation pressure and also to selection effects of the fast atoms through ionization processes. 

In that respect, we can argue that the decrease in LOS temperature between 0 to 50\deg\ is a clear signature
of the existence of two hydrogen populations contributing to the total line profile. Following Izmodenov et al. (2001),
we can divide the hydrogen atoms in the heliosphere into four distinct populations. First, we consider the interstellar
component that has gone through the heliospheric interface without any interaction with the protons. This population at large
distance from the sun has the distribution parameters of the interstellar gas, i.e. a bulk velocity close to 26 km/s and 
a temperature close to 6000K. The second population is the one created by charge exchange with protons of the compressed
interstellar plasma. Models predict a strong deceleration and heating of this population. These two populations are the main
contributors to the IP line profile at 1 AU, while the other two populations created by charge exchange with the solar wind 
can be neglected here (Qu\'emerais and Izmodenov 2002).  

\begin{figure}
\noindent\includegraphics[width=7.5cm]{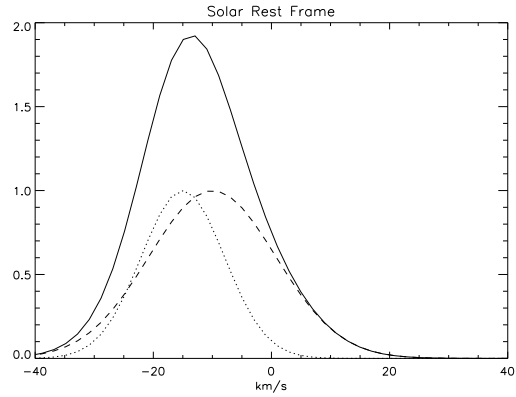}
\caption{ Model of the line profile generated for two distinct populations of hydrogen atoms. 
One has the parameters of the interstellar gas, i.e. velocity of 30 km/s 
(26 km/s accelerated by selection effects, see text) and a temperature of 6000 K, the other has 
parameters reflecting a deceleration and heating due to the heliospheric interface,  
i.e. velocity of 20 km/s and a temperature of 14000 K. The resulting line profile
in the upwind direction has a mean shift of -25 km/s and a temperature of a bit less than 
14000 K. The line profiles are shown in the solar rest frame.}
\label{model2pop}
\end{figure}

A simple model of the two populations is shown in Fig. \ref{model2pop}. The profiles are represented by Gaussian functions.
The plots are made in the solar rest frame. One population has the parameters of the interstellar gas, 
i.e. a velocity of 30 km/s and a temperature of 6000 K. We used a velocity of 30 km/s for the primary component to account 
for selection of faster atoms by ionization processes (See Qu\'emerais and Izmodenov 2002, Table 4).
The other component is decelerated in the solar rest frame and heated. Its parameters are given by a velocity of 20 km/s 
and a temperature of 14000 K. We have computed the line shift and line width
of the sum of these two line profiles projected on an LOS with an angle from upwind between zero and 50\deg . The results
are shown in Table \ref{mod2poptab}. We find values similar to the observations.

We do not claim that this simple model fits the data. It is just an example to illustrate
why the LOS temperature decreases between 0 and 50\deg , i.e. because the Doppler shift between the two 
components of the line decreases as the cosine of the angle with the upwind direction. After 50\deg\ from upwind, 
dynamic effects on the hydrogen distribution make the line width increase again. Actual modeling is required here
to correctly interpret these data.

\begin{table*}
\caption{ LOS velocity and temperature for a 2-population model \label{mod2poptab}} 
\begin{tabular}{c|c|c}
   LOS Angle from Upwind & LOS Velocity & LOS Temperature \\
   \hline
       0\deg\  &  -23.9 km/s    &  13733 K \\
      10\deg\  &   -23.6 km/s   &   13646 K \\
      20\deg\  &   -22.5 km/s   &   13394 K \\
      30\deg\  &   -20.7 km/s   &   13008 K \\
      40\deg\  &   -18.3 km/s   &   12535 K \\
      50\deg\  &   -15.4 km/s   &   12032 K \\
      60\deg\  &   -12.0 km/s   &   11559 K \\
\end{tabular}
\end{table*}

\begin{figure}
\noindent\includegraphics[width=7.5cm]{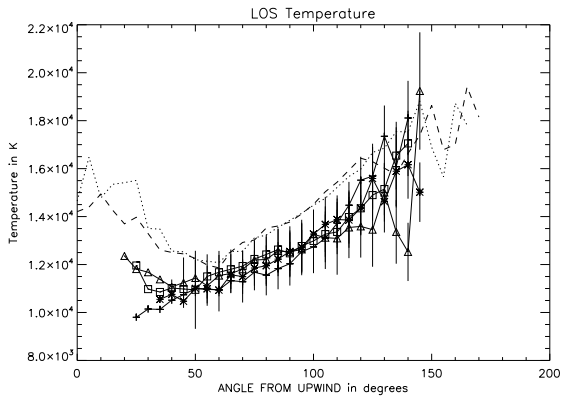}
\caption{ LOS Temperature as a function of the upwind angle. The values were obtained for the four orbits
from 1999 to 2002. The data from 1999 are shown by crosses, those of 2000 by stars, those of 2001 by squares,
and the data of 2002 are shown by triangles. There are no values for angles smaller than 
20\deg\ or larger than 150\deg\ because of contamination by stellar light.  The two curves from 1996 
(dashed line) and 1997 (dotted line) have been added for comparison. First, the temperature minimum seen 
previously around 60\deg\ seems to have shifted towards lower values. 
Second, the profiles appear cooler by roughly 1000 K  everywhere.  }
\label{temp9603}
\end{figure}

Figure \ref{temp9603} shows the LOS temperatures found for the four orbits from 1999 to 2002. The data from these
orbits are noisier due to the degradation of the sensitivity of the sensor units (Qu\'emerais and Bertaux 2002).
It was not possible to recover profiles for LOS's within 20\deg\ of the upwind direction or within
30\deg\ of the downwind direction. However, two results can be seen from Fig. \ref{temp9603}. First, the temperature
minimum around 60\deg\ from upwind has either shifted to lower angle values or even disappeared. This suggests that
the shift between the two components is smaller than in 1996 or that one of the components has become less important
relative to the other giving more weight to the parameters of the other component in the line profile. 
Second, the curves show lower temperatures than what was found in 1996 and 1997. The overall decrease is around
1000 K. This feature also suggests that one of the components has become relatively less important, thus yielding
an apparently cooler profile. It is unlikely that the shift between the components has decreased, because this
wouldn't change the LOS temperature for crosswind lines of sight.

The LOS temperature profiles presented in this section show that the IP line profile is most likely made up
of two components due to distinct hydrogen populations. The  first one is composed of unperturbed interstellar
hydrogen atoms getting close to the sun. The second one is composed of hydrogen atoms created after charge exchange with
interstellar protons compressed and heated in the heliospheric interface. The bulk velocity difference explains
why the IP line profile line width decreases when the upwind angle of the LOS goes from zero to 50\deg .
We also find that the profiles appear cooler during solar maximum than during solar minimum.
This could indicate that the increase of the ionization processes but also radiation pressure 
with the solar cycle is more effective on one population than the other. The slower population will be more ionized
than the fast one for instance. Consequently, this will result in narrower line profiles. These results will be compared
with model computations in a future work. 

\section{Comparison with HST profiles}

Direct measurements of the \lya\ line profile require a very good spectral resolution which is not
easily obtained with a space instrument. Fortunately, the STIS instrument on the Hubble Space 
Telescope has provided a few measurements in June 2000 and March 2001. This section presents the available
upwind spectrum and a comparison with the results obtained from the SWAN H cell data.

\subsection{The data}

The data presented in this section were obtained with the STIS instrument on-board the
Hubble Space Telescope.
One measurement was obtained in March 2001 (Upwind LOS).

The main problem for the observation of the interplanetary line profile from Earth's orbit is
caused by the existence of the strong emission of the geocorona. 
At the altitude of the Hubble Space Telescope, the geocoronal emission is 5 to 15 times brighter
than the interplanetary line depending on the direction of the LOS.
The best time to observe the interplanetary line is when the Doppler-shift between the two lines
is at its maximum. For the upwind direction, this is in March when the Earth velocity vector is toward 
the upwind direction. In that case, the relative motion between the H atoms and the observer is close to
50 km/s. 

%\begin{figure}[htb] 
%\noindent\includegraphics[width=15.0cm]{FIGURES/figexpliDoppler.ps}
%%\psfig{file=FIGURES/figexpliDoppler.ps,height=10.cm,width=15.0cm}
%\caption{\sl \label{fidoppler} Sketch showing the relative velocity between the interstellar gas and
%the Earth. The solar velocity (modulus=30 km/s) rotates around the Sun in one year. The LISM velocity
%vector (modulus=25km/s) is constant but slightly off the ecliptic plane. The relative velocity vector 
%which is the difference of the two previous vectors varies between roughly 10 km/s in september
%to 50 km/s in March. The best time for observation in a given direction is when the relative motion
%between the H atoms and the Earth projected on the LOS is the largest.}
%\end{figure}

Crosswind observations are limited to a relative velocity of 30 km/s because their LOS is
perpendicular to the velocity vector of the interstellar H atoms.
 This maximum value is obtained when the Earth
velocity vector is perpendicular to the interstellar wind direction, i.e. either when the 
Earth is upwind from the Sun (early June) or downwind (early December each year).

\begin{figure}[htb] 
\noindent\includegraphics[width=7.5cm]{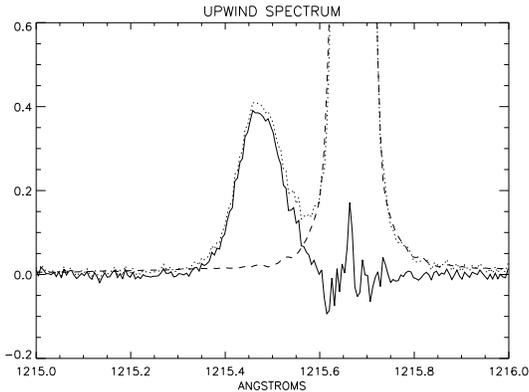}
\caption{ Upwind Interplanetary Spectrum measured by STIS on March, 21 2001. 
The geocorona (dash line) is more intense than the IP line (factor of 6) but the two lines are 
well separated because of the Doppler shift between the H atoms and the Earth. 
The geocoronal emission has been fitted and removed, leaving a larger uncertainty longward of the 
interplanetary line.}
\label{UpwindHST} 
\end{figure}

The upwind measurement was obtained on March 21, 2001 when the Doppler shift between the interplanetary 
line and the geocorona is close to its maximum (Position 2). In that case, the two lines are well separated. 
Figure \ref{UpwindHST} shows both the geocorona and the interplanetary line. A symmetric line shape fitting the 
geocorona has been removed from the data. The residual shows the upwind interplanetary line.

We estimated the LOS values of the velocity and the temperature of the line. The results for
apparent velocity corrected for the Earth's motion is 
\begin{center}
$V_{LOS} = -20.4 \pm 0.2$ km/s ,
\end{center}
the LOS temperature is 
\begin{center}
$T_{LOS} = 16 570 \pm 500$ K .
\end{center}
Those values are larger than the actual values because of the convolution of the actual line profile
with the line spread function of the instrument. The geocoronal emission from Fig. \ref{UpwindHST}
gives a good estimate of the line spread function (LSF). Its LOS temperature is equal to 5300 K whereas
the actual temperature of the geocorona is around 1000 K. 

\begin{figure}[htb] 
\noindent\includegraphics[width=7.5cm]{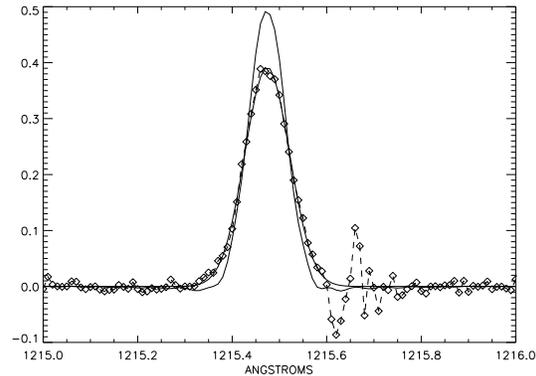}
\caption{ Upwind interplanetary spectrum in the solar rest frame.
The diamonds show the data, the thin solid line shows a Voigt function fit to the data,
the thick solid line shows the deconvolved spectrum obtained from the Voigt function fit
assuming that the LSF is given by the geocoronal profile.}
\label{UpwindDeconv}
\end{figure}

We have used the LSF deduced from the geocorona
to deconvolve the upwind line profile. The result is shown in Fig. \ref{UpwindDeconv}. 
First, the data were fitted to a Voigt function.
Then, this function was deconvolved yielding the spectrum. By assuming that the geocorona 
gives the LSF, we have slightly overestimated the actual width of the LSF, as reflected in 
the larger uncertainty in the temperature estimate. Future observations of the martian neutral 
H atom emission at \lya\
will better estimate the LSF because the martian emission profile has a thermal width 
equivalent to a few hundred K ($\approx$ 200 K). 

The resulting parameters for the upwind LOS are after deconvolution,
\begin{center}
$V_{LOS} = -20.3 \pm 0.2$ km/s .
\end{center}
The LOS temperature is 
\begin{center}
$T_{LOS} = 10 970 \pm 1000$ K .
\end{center}
The results found from the HST upwind line profile are compatible with the SWAN H cell measurements. 
First, the mean line shift measured in 2001 was 20.3 km/s. The SWAN result is 21.4 km/s. Taking 
a possible bias into
account due to the removal of the geocoronal line from the HST spectrum, we get a
correct agreement, thus confirming the change in the line shift of the IP line from solar minimum to
solar maximum.  Note also that a previous HST measurement made by GHRS (Clarke et al 1995, 1998) 
agreed with the SWAN value of -25.7 km/s for the solar minimum IP mean line shift.

The temperature found from the profile is close to 11000 K. This value may be slightly
underestimated because the LSF of the STIS instrument is not as wide as the Earth's coronal line. Comparing
with the values shown in Fig. \ref{temp9603}, we find correct agreement. This also confirms that
the upwind IP line profile LOS temperature has decreased from 14000K to 11000K from solar minimum 
to solar maximum It also shows that the temperature inflexion seen at 60\deg\ from upwind at solar minimum
more or less disappeared at solar maximum.

\section{Conclusion and  discussion}

This analysis of the SWAN H cell data has allowed us to reconstruct interplanetary \lya\ line profiles
between 1996 and 2003. This period covers the solar activity minimum of 1996 and the maximum of 2001.

We have found that the mean line shift changes from a LOS velocity of 25.7 km/s to 21.4 km/s in the solar
rest frame. This deceleration is mainly due to changes in radiation pressure with increasing activity, although
changes of the ionizing fluxes are also involved. Detailed modelling will be necessary to reproduce this large
variation in the mean line shift.

A comparison with a spectrum recorded by STIS on HST yields a good agreement. The STIS line mean shift corresponds
to an LOS velocity of 20.3 km/s which, given the uncertainties and possible biases involved in both analyses,
is quite acceptable.
We should also point out that the mean line shift change seen by SWAN is very rapid. We find a variation
 by 3 km/s between 1997 and 1999. This seems hard to explain only by changes in radiation pressure. 

The SWAN H cell data were also used to determine line widths (LOS temperature).
It is found that the line width variation with upwind angle is not monotonic as is usually found
from hot model computations. This can be explained as a proof that the IP line profile is made of two distinct
components scattered by populations with different bulk velocities and temperatures. These different components
have been theoretically predicted by models of the hydrogen interaction of the heliospheric interface. Here
we have an observable effect on the line width  which is created by these two populations. Actual model
computations of hydrogen distribution and backscattered profiles will be made to test this explanation.

Finally, we found that the LOS temperature profiles are cooler during solar maximum than solar minimum.
The line width change corresponds to a decrease in the LOS temperature by 1000 K.
A tentative explanation is that the slow hydrogen population component is more effectively ionized than the
fast one during solar maximum. This results in a smaller contribution to the total line shape and hence a narrower
line profile.

The results obtained in this analysis are summarized by this list

\begin{itemize}

\item
Between 1996 to 2001, the mean line shift of the interplanetary \lya\ line changes from a LOS velocity of 
25.7 km/s to 21.4 km/s in the solar rest frame.

\item
The line shift changes abruptly between 1997 and 1999 by 3 km/s.

\item
During solar minimum, the IP \lya\ line width shows a variation with the angle from upwind, which is not monotonic 
but has a minimum around 60\deg . This suggests that the IP line is composed of two components with different mean 
line shifts.

\item
During solar maximum, the LOS kinetic temperatures decrease slightly and the minimum is less pronounced,
suggesting that the ratio between the different components is changed from solar minimum conditions.

\item
Spectra obtained by HST in 1995 and 2001 give LOS velocities and temperatures are compatible
with the SWAN H cell measurements.

\end{itemize}

These empirical results will be confronted to model calculations in a future work.

\begin{acknowledgements}
SOHO is a mission of international cooperation between ESA and NASA.
SWAN was financed in France by the CNES with support from CNRS and in Finland
by TEKES and the Finnish Meteorological Institute.

%{\em Please provide any other acknowledgement information.}

\end{acknowledgements}

\end{document}